\documentclass[sn-basic,iicol]{sn-jnl}
\usepackage{siunitx}

\jyear{2022}%

\theoremstyle{thmstyleone}%
\theoremstyle{thmstyletwo}%

\theoremstyle{thmstylethree}%

\begin{document}

\title[Article Title]{Fully Automated Construction of Three-dimensional Finite Element Simulations from Optical Coherence Tomography}

\author[1,2]{\fnm{Ross} \sur{Straughan}}\email{ross.straughan@hest.ethz.ch}
\equalcont{These authors contributed equally to this work.}

\author[3]{\fnm{Karim} \sur{Kadry}}\email{kkadry@mit.edu}
\equalcont{These authors contributed equally to this work.}

\author[4]{\fnm{Sahil} \sur{A.Parikh}}\email{sap2196@cumc.columbia.edu}

\author[3,5]{\fnm{Elazer} \sur{R. Edelman}}\email{ere@mit.edu}

\author*[1]{\fnm{Farhad} \sur{R. Nezami}}\email{frikhtegarnezami@bwh.harvard.edu}

\affil*[1]{\orgdiv{Cardiac Surgery Division}, 
\orgname{Brigham and Women’s Hospital, Harvard Medical School}, 
\orgaddress{\city{Boston}, \postcode{02115}, \state{MA}, \country{USA}}}

\affil[2]{\orgdiv{Department of Mechanical and Process Engineering}, \orgname{ETH Zurich}, \orgaddress{\street{Leonhardstrasse 21}, \postcode{8092 Zurich}, \country{Switzerland}}}

\affil[3]{\orgdiv{Institute for
Medical Engineering and Science}, \orgname{Massachusetts Institute of Technology}, \orgaddress{\street{77 Massachusetts Ave}, \city{Cambridge}, \postcode{02139}, \state{MA}, \country{USA}}}

\affil[4]{\orgdiv{Cardiovascular Division}, 
\orgname{Division of Cardiology, Columbia University Irving Medical Center}, 
\orgaddress{\city{New York}, \postcode{10032}, \state{NY}, \country{USA}}}

\affil[5]{\orgdiv{Cardiovascular Division}, 
\orgname{Brigham and Women’s Hospital, Harvard Medical School}, 
\orgaddress{\city{Boston}, \postcode{02115}, \state{MA}, \country{USA}}}

\abstract{Despite recent advances in diagnosis and treatment, atherosclerotic coronary artery diseases remain a leading cause of death worldwide. Various imaging modalities and metrics can detect lesions and predict patients at risk; however, identifying unstable lesions is still difficult. Current techniques cannot fully capture the complex morphology-modulated mechanical responses that affect plaque stability, leading to catastrophic failure and mute the benefit of device and drug interventions. Finite Element (FE) simulations utilizing intravascular imaging OCT (Optical Coherence Tomography) are effective in defining physiological stress distributions. However, creating 3D FE simulations of coronary arteries from OCT images is challenging to fully automate given OCT frame sparsity, limited material contrast, and restricted penetration depth. 

To address such limitations, we developed an algorithmic approach to automatically produce 3D FE-ready digital twins from labeled OCT images. The 3D models are anatomically faithful and recapitulate mechanically relevant tissue lesion components, automatically producing morphologies structurally similar to manually constructed models whilst including more minute details. A mesh convergence study highlighted the ability to reach stress and strain convergence with average errors of just 5.9\% and 1.6\% respectively in comparison to FE models with approximately twice the number of elements in areas of refinement. Such an automated procedure will enable analysis of large clinical cohorts at a previously unattainable scale and opens the possibility for in-silico methods for patient specific diagnoses and treatment planning for coronary artery disease.} 



\maketitle

\clearpage

\section{Introduction}\label{sec1}

Cardiovascular diseases are the leading cause of death around the world \cite{RoMe19}. Atherosclerotic lesion stability is dependent on the balance of plaque structural stress (PSS) and material properties and composition, where plaque rupture is instigated once PSS exceeds material strength \cite{BrTe16}. Various studies have highlighted that easily determinable morphological markers such as the size of lipid core, lipid arc angle, fibrous cap thickness and lipid and calcium content can improve identification of unstable lesions \cite{KiLe12,GnHu17,BlMo17}. Despite suffering from reduced predictive power, current clinical metrics extracted from OCT imaging are widely adopted given their ease in extracting morphological features. However, quantitative metrics based merely on morphology alone often misidentify the risk of plaque rupture \cite{CaOb11,IdHu05,DoPa20}. In contrast, functional metrics informed by computational modelling of coronary micro-mechanics are more effective in identifying unstable lesions than those informed by morphology alone \cite{DoPa20}.

Patient-specific computational modeling by Finite Element Analysis (FEA) is a powerful technique to delineating the micromechanics of coronary artery, enabling the determinations of the aberrant mechanical stress distributions \cite{MiLe21}. However, many studies are limited to 2D geometries which facilitates ease of analysis but restricts accurate modeling \cite{HuVi01,MiLe21,GiVi20,BuEb14,NiAk13,OhFi05}. 2D FEA simulations tend to overestimate stresses within plaques \cite{NiAk13,OhFi05}, with some studies unrealistically estimating plaque stress to be as high as 1000 kPa  \cite{DoPa20}, whilst underestimating the stress concentrations produced by small calcifications \cite{KeMa13}. In contrast, 3D simulations are more accurate than 2D approaches but require extensive effort and laborious annotation time to create digital models, which limits scalability and repeatability.

Automatic segmentation of the atherosclerotic constituents is challenging in optical coherence tomography OCT images which suffer from increasing signal attenuation, resulting in limited penetration depth. Intravascular ultrasound (IVUS) better discerns arterial borders prompting hybrid modality approaches \citet{GuGi18} combining the high-resolution tissue components of OCT images with anatomical borders  obtained from IVUS  \cite{GuGi18}. However, clinical imaging is rarely performed with both IVUS and OCT, limiting the utility and feasibility of such an approach. Overcoming the shortfall of OCT imaging, \citet{OlAt18} fit a 3D anisotropic linear-elastic mesh to visible regions of the lumen and outer vessels, enabling fully automated delineation of the inner and outer borders. Neural network approaches by \citet{AtOl19} and others allowed for automated labelling of raw OCT into six categories: calcium, lipid tissue, fibrous tissue, mixed tissue and non-pathological tissue/ media \cite{OlAt18,OlNi22}. 

Even with effective methods to accurately segment the various tissues in OCT images, the conversion of labeled frames into digital twin computational models present several difficulties. Unlike many other  imaging modalities, OCT suffers from frame sparsity, often requiring manual 3D reconstruction followed by manual meshing \cite{CaGh22}. A fully automated pipeline that generates 3D FEA-ready digital twins from OCT pullbacks has not yet been created. \citet{Ka21} automated much of the process, with specialized techniques for meshing the 3D geometry automatically; however, the process still required extensive manual effort to create the 3D CAD model. An automated framework would significantly reduce the time taken to create 3D models with complex morphology by avoiding previously necessary manual extruding techniques to loft the material labels in each frame \cite{Ka21,GuGi18,AsHi13,WaPa20}, enabling structural simulations to be performed at an unprecedented scale, fully-capturing the patient-specific  mechanical response from complex morphologies \cite{Ka21,MiDo21,HoMu14}. We present a pipeline to automatically create 3D FE models that interpolates labeled OCT frames into a 3D multi-label voxelmap, meshes the 3D reconstruction with localized mesh refinement, and prescribes boundary conditions to the FE model, significantly accelerating the previously manual process. The framework we have developed presents a robust and highly repeatable procedure for transforming labeled OCT frames, with their inherent complex morphology, into three-dimensional reconstructions of coronary lesions through a series of processing steps that ensure a physiological and representative depiction of a patient's anatomy. This is followed by an optimized meshing technique that generates finite element models that generates physiological and consistent stress distributions. Our pipeline represents a significant advancement as it enables the automatic generation of digital twins for patients with CAD using raw intravascular images. This work allows us to calculate 3D structural stress within different lesion constituents, a capability that was not possible before. This opens new possibilities for studying lesion progression, predicting adverse clinical events, and aiding in therapy planning. By harnessing the power of digital twins, we can provide a more comprehensive understanding of CAD and improve patient-specific treatment strategies.

\section{Materials and Methods}\label{sec2}

\begin{figure*}[htbp]
\begin{center}
\includegraphics[width=0.95\textwidth]{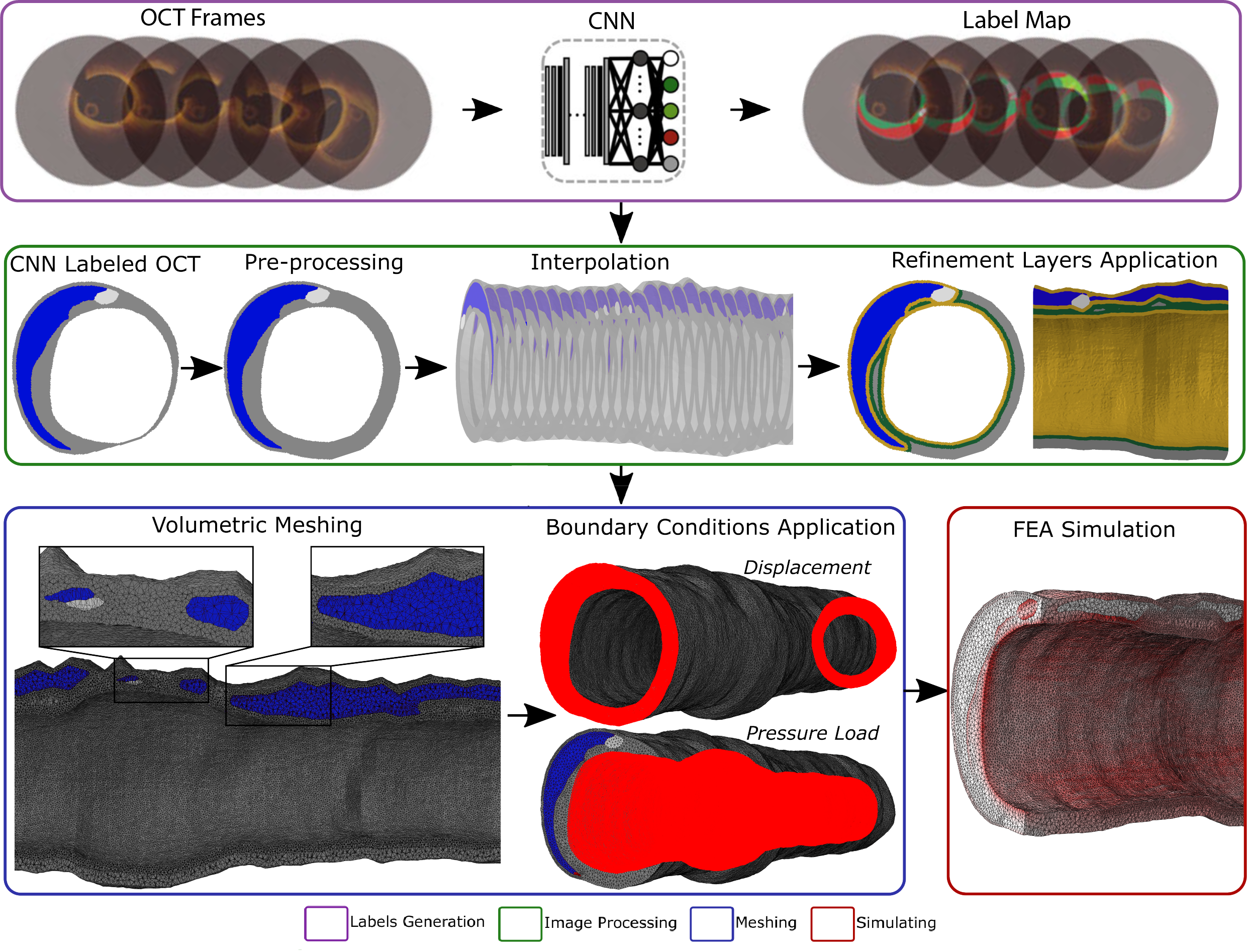}
\end{center}
\caption{The automated framework pipeline. A CNN developed by \citet{OlAt18} segments the raw OCT images into material labels corresponding to several tissue classes. Image processing steps modify the CNN labeled OCT frames to create anatomical geometry, followed by the signed distance function (SDF) based interpolation procedure. Healthy artery tissue, lipid and calcium are represented by the dark gray, blue, and light gray colors.. Refinement layers (gold and green) are then created to allow for extra refinement at sites of expected high stress. The golden layers are refined the most followed by the layer in green. Meshing steps utilizing open-source software create the volumetric mesh with the applied boundary conditions and material properties. The FE ABAQUS file is then ready to be used in a simulation.}\label{fig:flowchart}
\end{figure*}

Our framework (Fig. \ref{fig:flowchart}) took as input a 3D voxelgrid corresponding to a set of labeled OCT frames. Additional processing was performed on these segmentations before the frames were interpolated, creating an isotropic 3D voxelgrid of the coronary artery, subsequently all material properties and boundary conditions were prescribed in an ABAQUS finite element simulation file. All software utilized were open-source and coded in Python and Octave running as a single program in a Linux environment.

\subsection{Tissue Label Generation}

OCT images were acquired from patients who underwent invasive angiography and had one or more clinically identified significant stenoses. OCT images were collected according to clinical standards and approved by the Ethics Committee of the institution with all patients providing informed consent for the OCT images to be used for research purposes \citet{TeRe12,OlAt18}. The investigators used a C7-XR FD-OCT optical frequency domain intravascular imaging system alongside the DragonFly catheter (St. Jude Medical, Lightlab Imaging Inc., Westford, MA, USA). Motorized, automatic pullback was performed at 0.5 mm/s, with 20µm frame resolution and  400µm spacing. A predeveloped CNN with a validation accuracy of 96.05\%, developed by \citet{AtOl19} and trained on 700 OCT images from 28 patients, was utilized to generate the material labels for the four investigated OCT pullbacks. The CNN was trained on 480 K augmented patches from 22 of the patients, whilst the patches from the remaining 6 patients were reserved as validation for the CNN. The patched-based architecture begins with an input layer tailored for 32 x 32 pixel patches. This precedes an assembly of convolutional, batch normalization, and ReLU activation layers, sequentially followed by max pooling layers; each convolutional filters being 3 x 3 in size, increasing from 8 to 128 across each repetition. The last iteration utilizes an average pooling layer and is then proceeded by a fully connected layer, dropout layer, another fully connected layer, softmax layer and the ultimate classification layer for the output prediction. Through experimentation, 45 layers was deemed to be most optimal. The model is updated using Stochastic Gradient Descent and was found to be optimal to reduce training time. Labels were produced by medical experts per \citet{AtOl19} generating calcium, lipid, fibrous and mixed tissue domains. The CNN-labels were processed so that a thickness of 200 µm separated the lipid from the lumen to avoid rare non-physiological cases of the lumen direct contact with lipid  \citet{Ka21}.

\subsection{Pre-Processing}

The platform initially requires the labels to undergo a series of pre-processing steps, to ease the creation of the 3D finite element model and to rectify any CNN generated labels that are not anatomical. Firstly, we simplified the platform to only consider calcium, lipid and pool remaining fibrous and mixed tissue into one group. The lumen was prescribed its own unique label by isolating the second largest connected component of the image (with the largest connected component being the background). Specifying areas of the lumen allows for loads to be prescribed more easily to the inner surface of the artery as explained in more detail in section \ref{sec:BC}. Small tissue components less than 150 pixels (0.06 mm\textsuperscript{2}) in size were filtered out. Calcium components were excluded from the filtration step given their small size and sparsity. Occasionally, the CNN can produce walls that are non-anatomically thin, therefore such regions were thickened so that the entire arterial tissue has a minimum thickness of 500µm, the lower bound of what is anatomical \cite{FaFu00}; thickening the walls further would reduce stress concentration \cite{ChCh04}, reducing the apparent severity of the lesion.

\subsection{Frame Interpolation}

Unlike Computed Tomography (CT) or Magnetic Resonance Imaging (MRI), OCT suffers from frame sparsity as a result of the in-frame resolution being lower than the out-of-frame resolution. Due to the large gaps between frames, it is difficult to create a volumetric mesh before creating an isotropic pixel resolution. Our software framework first interpolated the CNN labeled OCT frames using a signed distance function (SDF) based algorithm  \cite{RaUd90}. The SDF describes the function value of some point, P, to the closest point of the implicit surface, S, generating a distance map between two labels in the frames, allowing for linear interpolation to deduce virtual labels within the physical space of the two labeled frames. The points on the surface satisfy the following relationship \cite{ZhWa21}:

\begin{equation}
	S = \{{x\in R^{3}}{\vert} f(x) = 0 \}
\end{equation}
The values of $f(x)$ refer to the signed distances. 

Each material component was interpolated individually to the corresponding component in the adjacent frame. In cases where a component is absent from one frame, an artificial label (one voxel in size), was added to the frame red lacking the component. This artificial label is added at the same x and y component as the centroid of the corresponding component where it is present in the previous frame. Eachlabel was expanded to fill any unlabeled areas after the interpolation procedure. As the labels are also expanded radially, the arterial cross-section was unphysiological. Therefore, a mask of the cross-section of the artery was interpolated in the same SDF procedure and then used to crop the expanded labels to a physiological cross-sectional area through a Boolean operation. The result of the process is an isotropic 3D reconstruction of the OCT image with smooth transitions between each interpolated frame (see Fig. S1).

\subsection{Post-Processing}

The mesh was locally refined near plaque components to allow for the efficient allocation of elements to the computational model, thereby budgeting the modelling cost. The vast majority of stress concentrations were expected to form in the vicinity of the atherosclerotic tissue components due to varying material stiffness or the inner wall of the artery based on the mechanics of pressurized cylinders. \cite{DoOt20,CaMa21}. Therefore, higher levels of mesh refinement were required in the surrounding arterial tissue to reach stress convergence. Two additional labels were thus constructed - an inner refinement layer applied to immediately surround the calcium, lipid and lumen, and an additional outer refinement layer surrounding the inner refinement layer to provide finer spatial control over the mesh generation (Fig.\ref{fig:flowchart}:Image Processing).

\subsection{Meshing}

An open-source mesh generating API, Iso2Mesh (ver 1.9.6), was utilized to create the initial meshes leveraging the CGAL (Computational Geometry Algorithms Library) algorithm (ver 5.0.2) due to its ability to mesh labelmaps with more than two labels intersecting (Fig. \ref{fig:flowchart}:Meshing) \cite{FaBo09,cgal}. A global mesh size was prescribed to all labels aside from the refinement layers with the inner layer having the finest mesh quality assigned. The resulting mesh consisted of both surface and volumetric elements. GMSH, another open-source meshing API, removed any 2D, surface elements so that only 3D, volumetric elements remained. Linear, C3D4 elements were then optionally converted to second order, C3D10 elements \cite{gmsh}. Additionally, the volumetric mesh was optimized using the Netgen algorithm to improve mesh quality and avoid associated mesh errors during FE simulation \cite{Sc97}. The final mesh was then converted to an \textit{inp} file, a native ABAQUS finite element input for structural simulations. When creating the inp file, the C3D4 or C3D10 elements were converted to their respective hybrid equivalents (C3D4H and C3D10H). Information regarding the connectivity of nodes associated with the lumen was extracted (discussed further in section \ref{sec:BC}), and subsequently the lumen was removed from the mesh.

\subsection{Boundary Conditions}\label{sec:BC}

Inner arterial wall nodes were defined by identifying nodes that were shared by any of the arterial components and lumen. Boundary conditions are applied by editing the \textit{inp} file (Fig. \ref{fig:flowchart}:Meshing). A pressure of 15 kPa was then applied to elements of the inner arterial wall to represent the physiological blood pressure of coronary arteries  \cite{BaHe19,Ka21}. Nodal boundary conditions were applied on both ends of the coronary artery to prevent displacement in all directions. Nodes associated with the end caps were identified by finding nodes within a specific distance from the maximum and minimum z-coordinates of each end cap.

\subsection{Material Properties}

Mimicking the isotropic model used by \citet{Ka21}, soft tissues were modeled as hyperelastic materials. A third-order polynomial strain energy function is described as:

\begin{equation}    
\begin{split}
		\psi = C_{10}(I_{1} - 3) + C_{01}(I_2 - 3) + C_{11}(I_{1} -3)(I_2 - 3)\\ + C_{30}(I_{1} - 3)^{3} + \frac{1}{D}(J - 1)^{2}  \label{eq:isotropic}
\end{split}
\end{equation}

\noindent Where the constants  Cij and D refer to the distortional and compressional response respectively. Constants $I_1$, $I_2$ and $I_3$ are the invariants of the right Cauchy-Green deformation tensor. The values for these constants (table \ref{table:isotropic}) were produced experimentally from ex-vivo mechanical experiments  \cite{KaNa13,MaTa19}.

Calcium, simply assumed to be a linear elastic material, was modeled using \citet{Ka21}:

\begin{equation}
	\sigma = E \epsilon  \label{eq:calc}
\end{equation}

\noindent Where E refers to the young's modulus with a value of 184 MPa \cite{EbCo09} based on experimental data with a Poisson's ratio of 0.495, as assumed by \citet{Ka21}.

\begin{table}[htb!]
	\centering
	\begin{tabular}{lllllll}
		\multicolumn{1}{p{0.35cm}}{\centering C10 \\ {[kPa]}} &
		\multicolumn{1}{p{0.35cm}}{\centering C01 \\ {[kPa]}} &
		\multicolumn{1}{p{0.35cm}}{\centering C20 \\ {[kPa]}} &
		\multicolumn{1}{p{0.35cm}}{\centering C11 \\ {[kPa]}} &
		\multicolumn{1}{p{0.35cm}}{\centering C30 \\ {[kPa]}} &
		D     \\ 
		\cline{1-7}
		Healthy Artery & 
		127.9 & 
		0 & 
		0 & 
		0 & 
		0 & 
		0.096 \\
		Lipid & 
		1.6 & 
		0  & 
		9.3 & 
		0 & 
		11 &
		0     \\
	\end{tabular}
	\caption[Isotropic hyperelastic properties] {Hyperelastic constants used in the FE model for  healthy artery and lipid tissue.}
	\label{table:isotropic}
\end{table}

\subsection{FEA Convergence Experiment}

\begin{figure}[htb!]
\begin{center}
\includegraphics[width=0.475\textwidth]{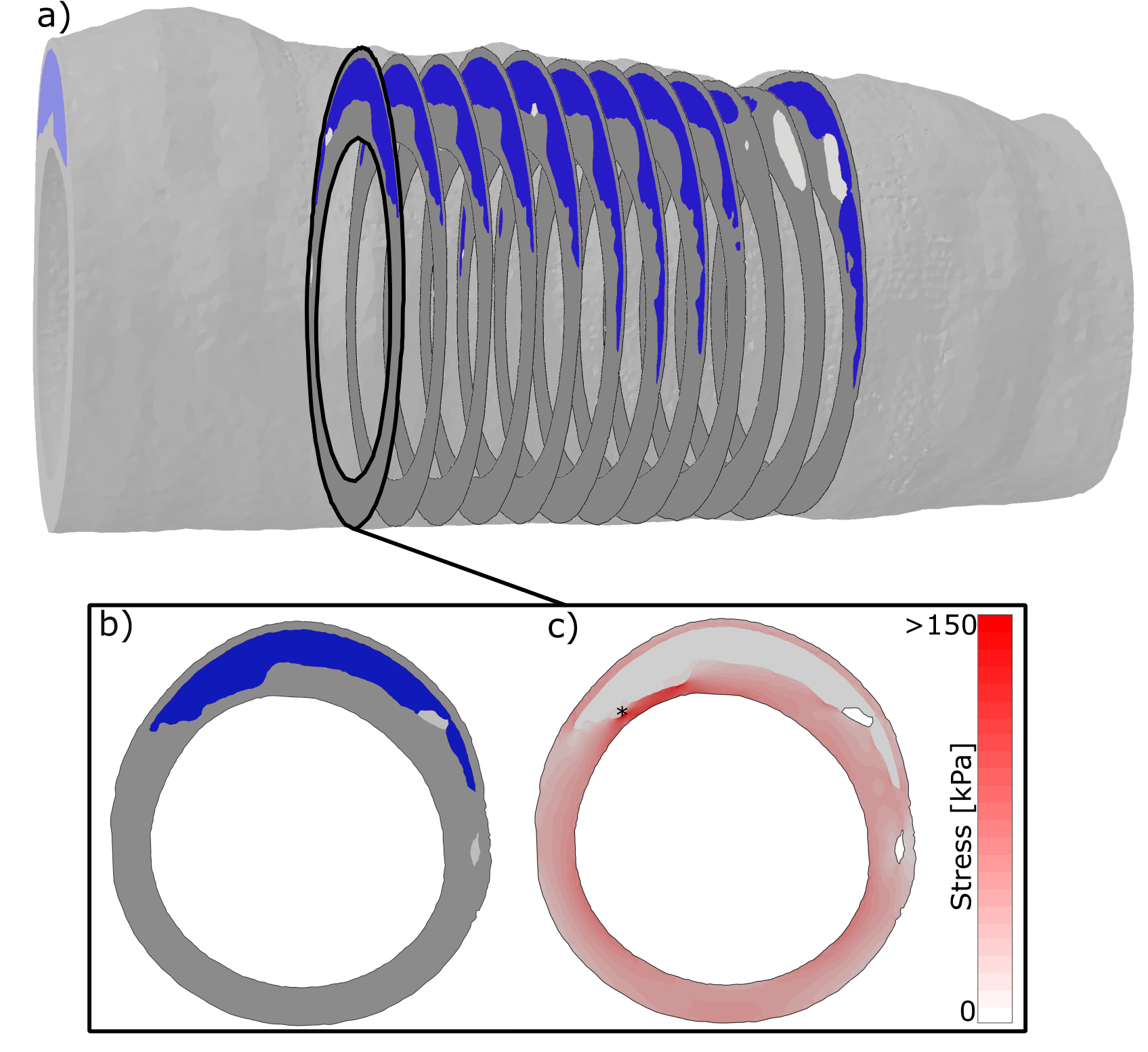}
\end{center}
\caption{Illustration of the frames that were investigated for the mesh convergence study. Frames shown in (a) consist of complex geometry that includes large contents of lipid (blue) and calcium (light gray) embedded within arterial tissue (dark gray). Region highlighted shows the material distribution (b) and tissue stress (c). The arterial stress is significantly concentrated at the region of the thinnest fibrous cap. Stress of calcium was excluded to highlight arterial stress distribution. Peak stress is highlighted by the (*) marker}\label{fig:mc_selection}
\end{figure}

A region of interest, 10mm in length, consisting of 25 physical OCT frames was selected for a mesh convergence experiment. A total of 8 separate FE models of this region of interest with various degrees of mesh refinement for the inner, most refined layer were investigated to test the capability of the framework in converging for peak strain and peak stress. Iso2Mesh generates elements of varying size by defining the maximum pixel volume size rather than the length of each element. The largest and smallest volume size was set at 2.83 and 1.72 cubic voxels  (with a voxel resolution of 20 µm), resulting in a near quadrupling of the number of refined elements from 375,000 elements to 1,374,000 elements. The global number of elements ranged from 1,052,000 to 2,257,000 for the least and most refined models respectively. A total of 12 2D OCT frames from this region of interest were selected for peak stress analysis. These frames were 2.8 mm away from the end caps of the artery to avoid any effects that the displacement boundary conditions may have on the final results. The selected frames (Fig. \ref{fig:mc_selection}) contained varied, complex geometry consisting of healthy arterial tissue, lipid and calcium. The refinement layer in the least refined mesh contained $3.75 \times \num{10}^{5}$ elements, whilst the most refined mesh contained $1.37 \times \num{10}^{6}$ elements.

\section{Results}\label{sec2}

\subsection{3D Reconstruction}

\begin{figure*}[ht!]
\begin{center}
\includegraphics[width=0.8\textwidth]{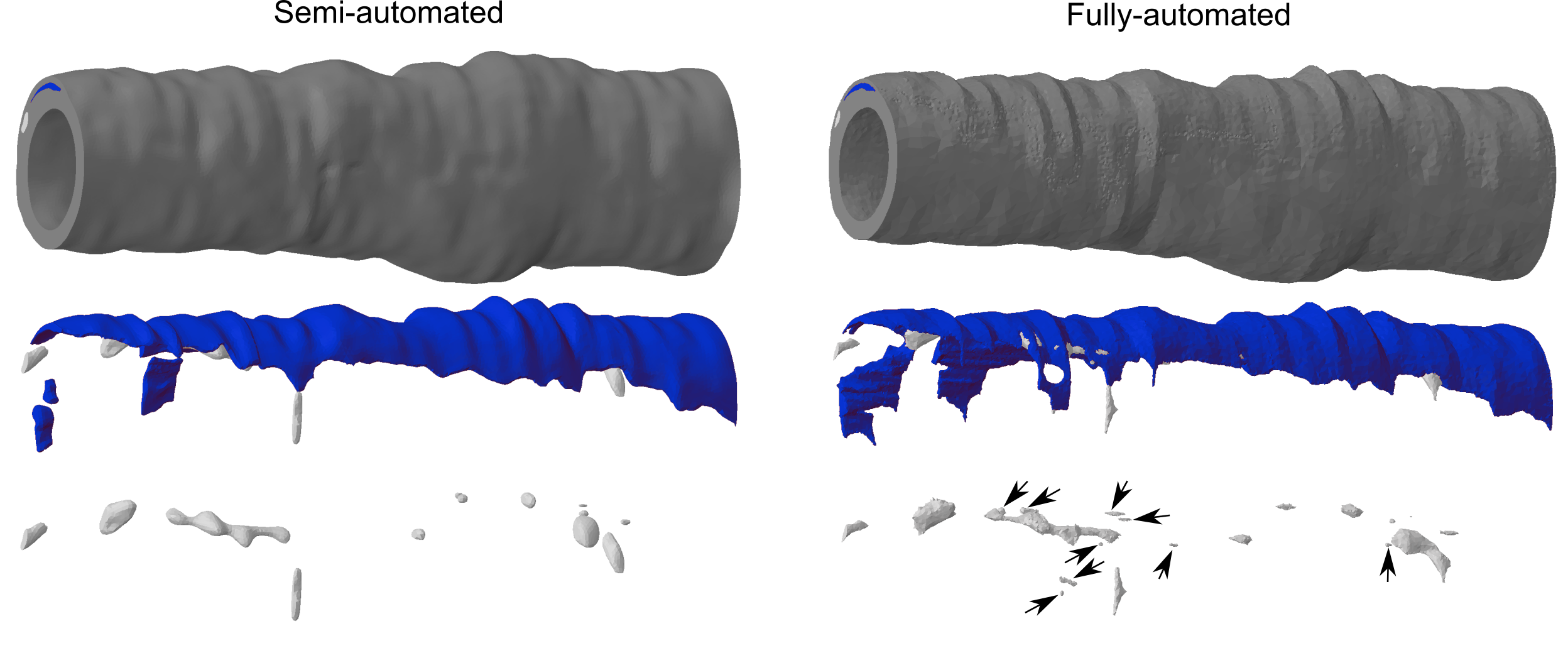}
\end{center}
\caption{Comparison of the generated 3D geometry created by the semi-automated CAD based method employed by \citet{Ka21} (left) vs the newly proposed method in this paper (right). The semi-automated method, that relied on manual lofting using CAD software, produces a model of similar morphology for healthy arterial tissue (dark gray), lipid (blue) and calcium(light gray). A noticeable difference is that the automated method includes some extra smaller, but mechanically significant calcifications (black arrows), otherwise unfeasible to include in the manual process.}\label{fig:3D_comparison}
\end{figure*}

It can visually be confirmed that the automatically generated 3D reconstruction in this paper produces an anatomical morphology (Fig. \ref{fig:3D_comparison}). When visually comparing the reconstruction produced by \cite{Ka21}, the larger morphological features are similar; however, the fully-automated method included more minute details such as smaller islands of calcification. Another noticeable difference is that the automated method tended to connect material constituents more frequently in the space between the labeled OCT frames than the semi-automated method.

\subsection{FEA Convergence}

Convergence is typically defined based on the percentage error of a measured value with the doubling of the number of elements.  Hence, a comparison was made between the most refined mesh (1.37 × 10$^{6}$ elements) and a mesh with approximately half the number of elements (6.79 ×10$^{5}$ elements). The degree of strain and stress error is a measure of the error with respect to the most refined mesh. Comparing these two meshes with varying degrees of refinement there was an average strain error of 1.61\% for each frame investigated and the error remained below 10\% for each individual frame. The average percentage error of the stress was 5.9\% with 9/12 frames having an error below 10\% and the remaining 3 frames below 15\% error.

\subsection{Stress Distribution}

\begin{figure*}[ht!]
\begin{center}
\includegraphics[width=0.95\textwidth]{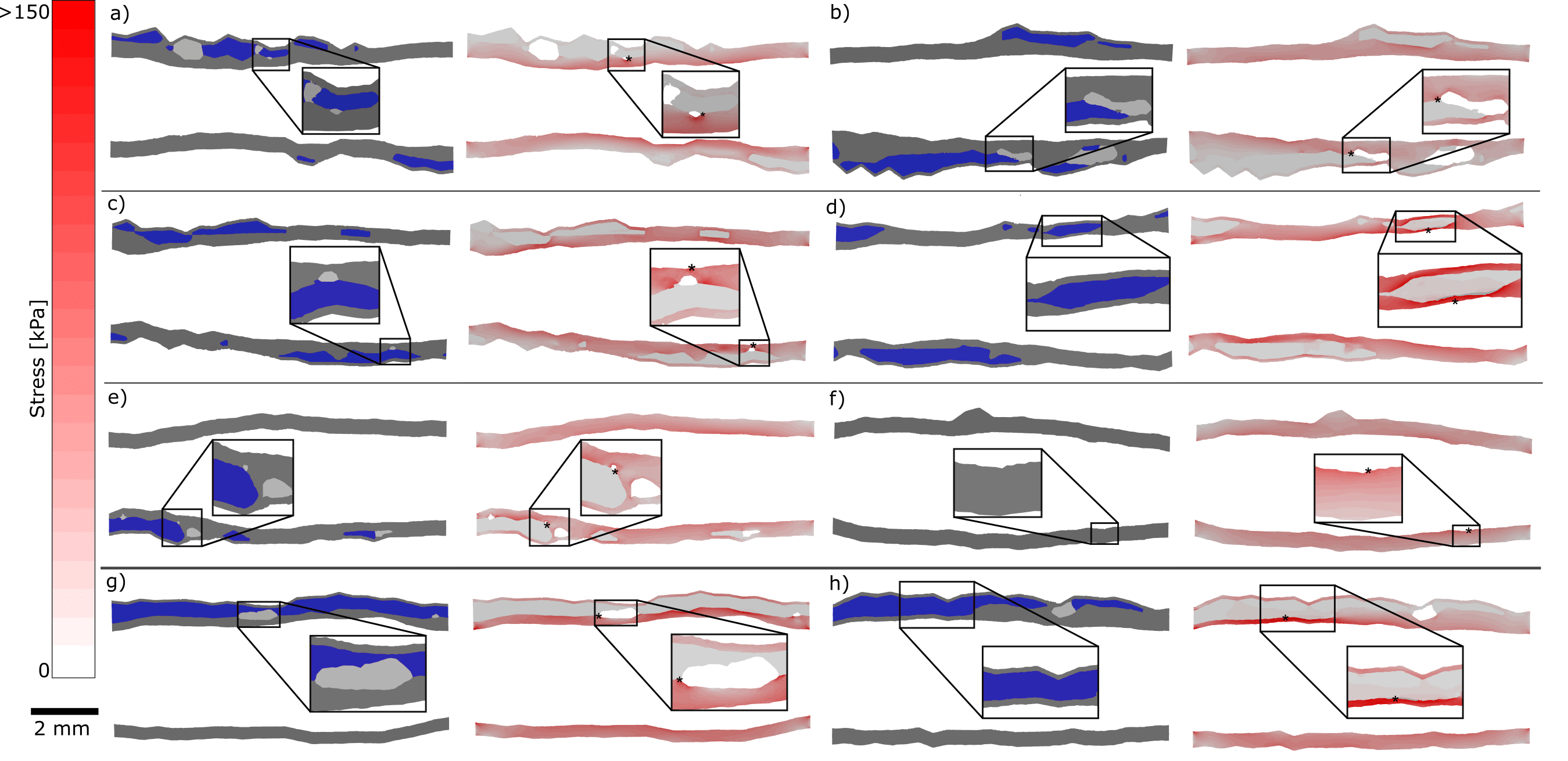}
\end{center}
\caption{Cross-sectional view of eight different sections from four separate OCT pullbacks, each exhibiting a distinct morphology, resulting in a variety of stress responses. The peak stress is represented by a (*) marker.}\label{fig:cut_view}
\end{figure*}

Several regions of interest were extracted from four separate OCT pullbacks to demonstrate the utility of developed framework to create 3D FEA-ready cases  (Fig. \ref{fig:cut_view}). We observed that peak stress was distributed near areas of high lipid content (Fig. \ref{fig:cut_view} (d), (g) and (h)), often located at lipid-calcium interfaces (Fig. \ref{fig:cut_view} (b), (e), and (g)) and were sharpest near calcium segments acting as localized stress inducers (Fig.  \ref{fig:cut_view} (b), (c), (e) and (g)). Small calcifications induced localized areas of high stress (Fig.  \ref{fig:cut_view} (a), (c) and (e)).

\begin{figure}[ht!]
\begin{center}
\includegraphics[width=0.475\textwidth]{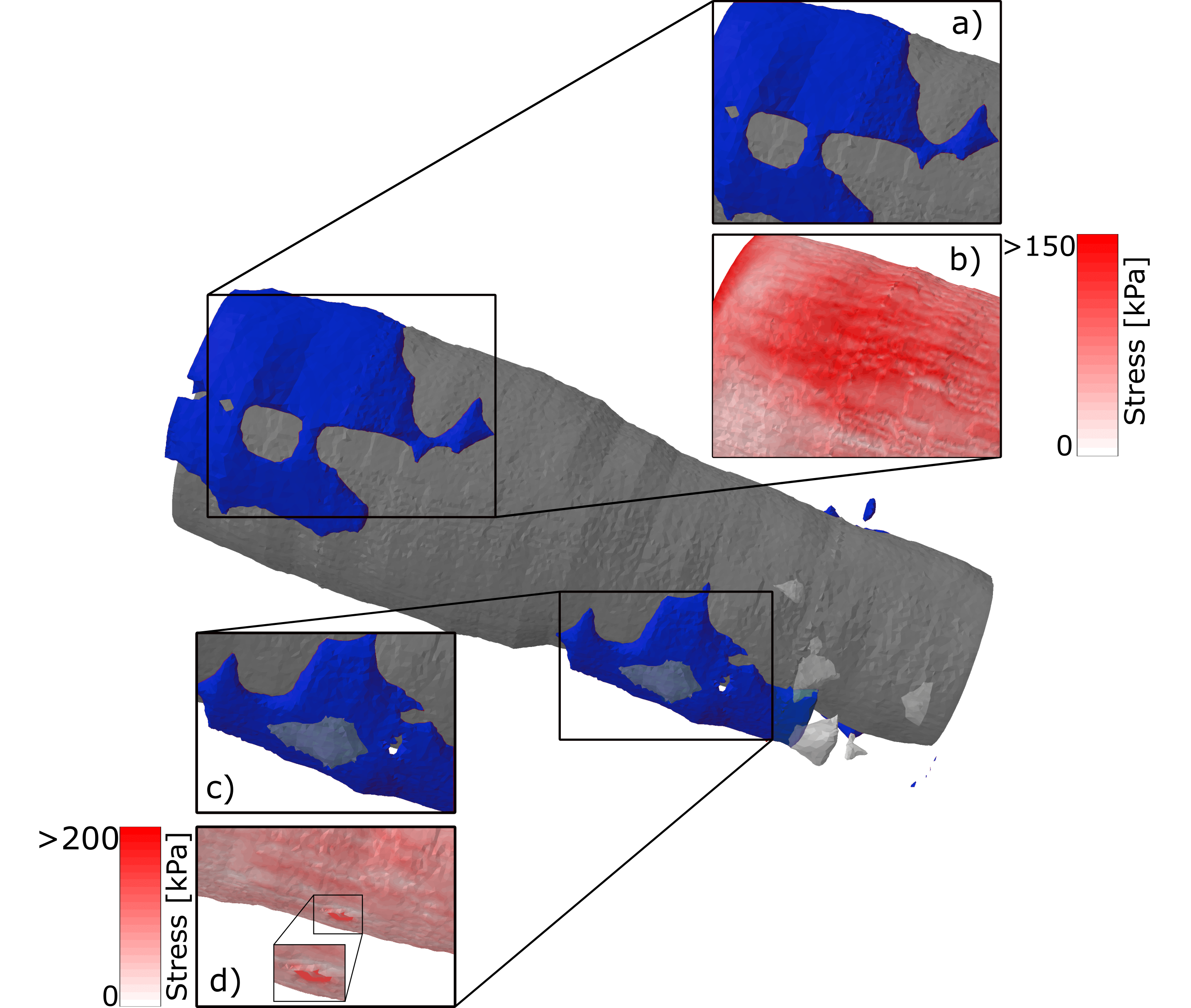}
\end{center}
\caption{View of the inner vessel wall surface of healthy arterial tissue with embedded lipid and calcium volumes. The positioning of the lipid and calcium volumes significantly impacts the surface stress of the artery. The large deposit of lipid that resides near the surface of the lumen (a), induces a region of higher stress concentration on the arterial surface (b). Although the volume of clarifications may be small in region (c), the stress inducing effects of a stiff localized deposit are significant (d). }\label{fig:surface_stress}
\end{figure}

In many of these cases, stress concentrations occurred at the inner surface of the arterial tissue (Fig. \ref{fig:cut_view} (a), (c), (d), (f), (g) and (h)), mainly due to large deposits of lipid nearby. This phenomenon is exemplified in Fig. (\ref{fig:surface_stress}) wherein (a) and (b) highlight a region of high lipid content with a small fibrous cap thickness, resulting in a region of high surface stress. Calcium is also capable of inducing arterial stress concentrations (Fig. \ref{fig:surface_stress} (c) and (d)) where calcification in direct contract with the lumen, a calcified nodule, induces a sharp increase and highly localized region of stress concentration.

\section{Discussion}\label{sec12}

OCT-based structural simulations are a useful tool in capturing patient-specific risk of atherosclerotic events. However, manually generating simulation ready models is cumbersome, limiting their utility and application. Such preliminary hurdles are inevitable due to complexity of generating 3D reconstructions from frames, difficulties with meshing the complex 3D geometry and convergence issues. A fully automated pipeline will significantly reduce the time and effort necessary to create structural models from OCT images. We, herein, introduce the first fully automated process to create 3D structural finite element models from OCT imaging by creating a pipeline that interpolates CNN-generated labeled OCT frames, meshes the generated 3D reconstructions and prescribes the necessary material properties and boundary conditions. The introduced framework enables an unprecedented level of scalability in generating structural simulations of atherosclerosis, in a highly-repeatable manner. Paired with fully automated segmentation techniques \cite{OlAt18}, the framework enables automatic calculation of mechanically informed metrics from an OCT imaging pullback, significantly boosting the clinical translatability of structural simulations to determine patient-specific risk.

Our framework creates 3D reconstructions of arteries from sparse OCT images, producing similar results to labor-intensive methods, whilst incorporating minute details such as small calcifications. Although automating much of the process, the lofting process of \citet{Ka21} could take up to 5 days depending on the complexity of the micromorphology of the lesion.  The meshing process of this prior method involved timely and computationally expensive boolean operations to remove overlapping volumes of the 3D CAD model, taking up to an hour to complete. Meanwhile, with our newly proposed method, the creation of the 3D FEA model from labelled data consistently required less than 20 minutes for the same investigated OCT pullback. Interpolation will inevitably lead to morphologies more elaborate and different from  manual attempts, yet the automated method closely matches an anatomical human-made method (see Fig. \ref{fig:3D_comparison}) demonstrating the reliability of the technique. Nevertheless, it has been shown that a reduction in the number of axial slices has minimal impact on the result on the general 3D stress distributions from FE simulations  \cite{NiAk13}. This may suggest that the method of 3D interpolation, provided that it is anatomical, may not significantly impact the stress distribution.

Although not based on OCT imaging, \citet{WaYo22} produced a fully automated procedure to create FE files based on IVUS imaging. Unassigned elements were initially created and then each element was assigned a material based on the closest label in a physical IVUS frame. In their technique, only surface meshes of the profiles of the lumen and outer arterial wall were generated, ignoring the surface of each individual material component; hence, the elements were not generated to represent the complex material morphology, yielding sharp and irregular interfaces between the various materials, likely to induce unphysiological stress concentrations. In contrast, the CGAL algorithm that we employ creates smooth surface meshes around each individual material constituent  (see Fig. \ref{fig:flowchart}), faithfully reproducing the morphology of the constituent materials of the artery. Peak arterial wall stress is often used as a quantitative measure of lesion stability  \citet{GuMa21,DoPa20}, elevating this metric in importance and amplifying any inaccuracies, limiting clinical utility. As the refinement layers in the method we now present are applied around areas of expected high stress nonphysiologic stress singularities and exaggerated instability are avoided.

There are a number of limitations with the current approach. We have not accounted for arterial curvature which may have a minor impact on the mechanical response of the tissue. However, this can be rectified by including centerline or catheter path information acquired from supplementary modalities such as Coronary Computed Tomography Angiography (CCTA) \cite{KaKa22}. Incorporating the anatomical curvature is more critical in fluid-structure interaction (FSI) models due to its impact on disturbing blood flow, leading to low and oscillatory wall shear stress \cite{LvWa22,WaYo22}. Our approach can incorporate lumen segmentation to create FSI models that can calculate both the mechanical and fluidic response of the lesion. Although PSS is likely to be a better predictor of short term mechanical stability, fluid dynamics, and wall shear stress (WSS) in particular, are believed to be mechanical cues for long term remodeling  \cite{SaEs11,GiKa19}. Utilizing both PSS  and WSS will enable a tool to predict both short-term and long-term prognosis of patients. 

Although the scope of this paper does not focus on techniques used to segment the OCT frames, the results of the structural simulations are highly dependent on the accuracy of the segmentation technique. Hence, incorporating state of the art labeling, image reconstruction and post-processing techniques will ultimately provide a greater representation of the mechanical state of specific lesions \cite{NiOl21,KaKa22, CaGh22, ElKa21, KaSa21, KaOl22}. Our method utilized CNN based segmentations to create labeled data, the accuracy of which is contingent on dataset size and variety of both the OCT imaging systems and that of the diseased and healthy coronary arteries represented within the dataset \cite{OlNi22}. Nonetheless, the scope of this paper focuses on utilizes utilizing labeled frames developed in any method and owing to the modular nature of the pipeline, the labeling method can conveniently be improved and adjusted to ultimately yield structural simulations of higher fidelity.

Unstable lesions are often identified when focusing on peak values of stress and strain in finite element analyses, but such studies rarely discuss and verify convergence of peak values. In contrast, our model produced converged results for strain even in complex lesions and with minimal mesh refinement. Although, the error of stress values was higher than strain values, the error was consistently below 15\% with a near doubling of the refined layer. Moreover, the three cases exhibiting more than 10\% error were all clinically less significant as their frames had the smallest stress peaks. Stress peaks for the majority of these cases occurred at various locations for each level of refinement, inevitably contributing to the large fluctuations in the recorded stress values. On the contrary, slices with more distinct, larger stress concentrations had stress peaks persistently at the same location for each level of refinement. Achieving mesh refinement demands the utilization of regional mesh refinement, especially for materials with highly varying mechanical responses, without so, determining lesion stability based on peak stress values will be unreliable. Rather than uniform element sizing in each refinement layer, a more efficient allocation of elements would increase element sizes gradually from high- stress areas. Additionally, a stress-adaptive proce- dure could aid in reaching convergence by seeding extra elements at high-stress sites after an initial simulation.
 
Including residual stress enables a more equal distribution of stresses throughout the wall of the arterial layers and therefore neglecting this effect in our framework may slightly misrepresent the physiological state of the artery \cite{GiSp22}. Computational modeling of residual stress for in-vivo based models is actively being researched and implementation into such a model may provide more accurate determination of stress distributions \cite{GiSp22,ScJo14,SiTa19}.

Accumulating evidence suggests that purely morphological assessments of intravascular imaging such as OCT have limited predictive capability in determining lesion stability \cite{GuMa21,DoPa20}. Fully automated structural analysis allow for mechanically informed lesion stability analysis in a clinical environment, enabling improved therapeutic planning. Virtual interventions can be performed to optimize device design and intravascular placement. Automated methods can lead the way for expansive trials across permutations and combinations of lesions and interventions and allow for mechanically informed metrics to assess plaque stability to emerge, augmenting current clinical guidelines.

\section{Conclusion}\label{sec13}

We present one of the first algorithmic approaches to automatically construct 3D finite element models from raw OCT imagery. Our approach recapitulates the 3D and multi-material nature of atherosclerosis, and predicts the mechanical response of lesions with a high degree of fidelity. The time required to create such patient-specific physics simulations have been reduced to a matter of minutes, significantly improving the ability to perform large scale research studies in addition to improving the clinical viability of the research field.

\backmatter



\bmhead{Acknowledgments}

Financial support was provided in part by the National Institute of Health (1R01HL161069) to ERE and FRN as well as the Zeno Karl Schindler Foundation to RS.

\section*{Declarations}
We declare there are no conflicts of interest.





\bibliography{sn-bibliography}


\end{document}